\documentclass[aps,prl,floatfix,twocolumn,preprintnumbers,amsmath,amssymb,groupedaddress]{revtex4}

\usepackage{graphicx}  
\usepackage{dcolumn}   
\usepackage{bm}        
\usepackage{stmaryrd}  
\usepackage{multirow}  
\usepackage{color}     
\usepackage{rotating}  
\usepackage{bbm}       

\begin{document}
\title{Band-structure topologies of graphene: spin-orbit coupling effects from first principles}
\author{M. Gmitra$^1$, S. Konschuh$^1$, C. Ertler$^1$,
C. Ambrosch-Draxl$^2$ and J. Fabian$^1$}
\affiliation{
$^1$Institute for Theoretical Physics, University of Regensburg,
93040 Regensburg, Germany\\
$^2$Chair or Atomistic Modeling and Design of Materials, University of Leoben,
Franz-Josef-Strasse 18, A-8700 Leoben, Austria
}


\begin{abstract}
The electronic band structure of graphene in the presence of spin-orbit coupling
and transverse electric field is investigated from first principles using the
linearized augmented plane-wave method. The
spin-orbit coupling opens a gap at the $K(K')$-point of the magnitude of 24 $\mu$eV (0.28 K).
This intrinsic splitting comes 96\% from the usually neglected $d$ and higher orbitals.
The electric field induces an additional (extrinsic)
Bychkov-Rashba-type splitting of $10$ $\mu$eV (0.11 K) per V/nm, coming from the $\sigma$-$\pi$
mixing. A `mini-ripple' configuration with every other atom is shifted
out of the sheet by less than 1\% differs little from the intrinsic case.
\end{abstract}
\maketitle


The fascination with graphene \cite{Novoselov2004:Science}, the one-atom-thick allotrope of carbon,
comes from its two-dimensional structure as well as from
its unique electronic properties
\cite{Novoselov2005:Nature,Zhang2005:Nature, Bostwick2007:NP,Geim2007:NM, Adam2007:PNAS}.
The latter originate from the specific
electronic band structure at the Fermi level: electrons move with a constant
velocity, apparently without mass and a spectral gap.
Analogy with massless Dirac fermions is often drawn, presenting graphene
as a solid-state toy for relativistic quantum mechanics. Ironically, this
nice analogy is broken by the relativistic effects themselves. In particular,
the interaction of the orbital and spin degrees of freedom, spin-orbit coupling,
gives the electrons in graphene a finite mass and induces a gap in the spectrum.
How large is the gap and which orbital states contribute to it? This question
is crucial for knowing graphene's band-structure topology, understanding its
spin transport and spin relaxation properties~\cite{Tombros2007:Nature, Kane2005:PRL},
or for assessing prospects of graphene for spin-based quantum computing \cite{Trauzettel2007:NP}.
By performing comprehensive first-principles calculations we predict the
spectral gap and establish the relevant electronic spectrum of graphene
in the presence of external transverse electric field.
We find that realistic electric fields can tune among different
band structure topologies with important ramifications
for the physics of graphene.

Carbon atoms in graphene are arranged in a honeycomb lattice which comprises
two triangular Bravais lattices; the unit cell  has two atoms.
The corresponding reciprocal lattice is again honeycomb,
with two nonequivalent vertices $K$ and $K'$ which are the Fermi momenta
of a neutral graphene. The states relevant for transport are concentrated in two touching
cones with the tips at $K(K')$---the Dirac points---as illustrated in Fig. 1. The corresponding Bloch
states are formed mainly by the carbon valence $p_z$ orbitals (the z-axis is
perpendicular to the graphene plane) forming the two $\pi$ bands (cones). The other three occupied valence states of carbon form the deep-lying $\sigma$ bands by $sp^2$ hybridization; these are responsible for the robustness of graphene's structure. The states in the lower cones are hole or valence like, the upper cone states are electron or conduction like, borrowing from semiconductor terminology. These essentials of the electronic band structure of graphene were worked out many decades ago \cite{Wallace1947:PR,Coulson1952:PPS,Lomer1955:PRS,McClure1957:PR,Slonczewski1958:PR,McClure1962}.

The above picture breaks down when spin-orbit coupling is included. The most important
modification to the band structure is the opening of a gap at $K(K')$, as predicted
by Slonczewski and Weiss \cite{Slonczewski1958:PR,Slonczewski:thesis}. More
severe changes occur when graphene is subject to transverse electric field that
can come from the substrate or electric gates. Several questions arise:
(i) What is the gap caused by the spin-orbit coupling itself? The magnitude
of this intrinsic gap gives the temperature scale for observing the spin-orbit effects.
While the spin-orbit coupling splitting of the $p$-orbitals in carbon atoms
is of order 10 meV, the relevant states in graphene comprise $p_z$ orbitals
which have no net orbital momentum along $z$ so the effect is expected to be rather weak. (ii)
For realistic electric fields, what are the corresponding extrinsic effects
on the band structure? Are the effects comparable to the intrinsic ones? The
answer helps to decide on the spin transport mode as well as on spin relaxation
and spin coherence mechanisms, or to see if the band structure can be
tailored. (iii) Which orbitals are involved in the intrinsic
and extrinsic effects? Finally, (iv) do transverse deformations of the graphene plane
induce significant spin-orbit effects?

To investigate the electronic band structure of graphene in the presence
of spin-orbit coupling and to answer questions (i) to (iv) we
employed the full potential
linearized augmented plane waves (LAPW) method based on density functional theory \cite{Blaha:Wien2k}.
For exchange-correlation effects we used the generalized
gradient approximation (GGA) \cite{Perdew1996:PRL}. In our three-dimensional
calculation the graphene sheets of lattice constant $a=1.42\sqrt{3}\,{\rm \AA}$
are separated by the distance $20\,{\rm \AA}$, large enough for the inter-sheet
tunneling to be negligible. Integration in the reciprocal space was performed
by the modified Bl\"ochl tetrahedron scheme, taking the mesh of
$33\times 33$ $k$-points in the irreducible Brillouin zone wedge.
As the plane-wave cut-off we took $9.87\,{\rm\AA^{-1}}$. The 1s core states were 
obtained by solving the Dirac equation, while spin-orbit coupling for the valence 
electrons was treated within the muffin-tin radius of 1.34 a.u.
by the second variational method \cite{Singh2005:book}. Finally, external 
transverse electric field was included as a periodic zigzag electric potential 
added to the exchange-correlation functional \cite{Stahl2001:PRB}.

Our main results are shown in Fig.~\ref{Fig:bands} which displays
a variety of the band structure topologies of graphene, tunable by electric
field. The intrinsic case (zero external field) shows a splitting of
the Dirac cones at the $K(K')$ points into two ``rounded'' cones, with the
gap of $2\lambda_I= 24$ $\mu$eV (0.28 K). The bands are two-fold degenerate
due to the presence of time reversal and space inversion symmetries \cite{Fabian2007:APS}.
Transverse electric field breaks the latter symmetry, resulting in
a spin-splitting $2\lambda_{\rm BR}$ of the energy levels: for each momentum
at a given band there are two states with energy differing by $2\lambda_{\rm BR}$.
This extrinsic splitting is akin to the Bychkov-Rashba (BR) spin-orbit coupling in
semiconductor heterostructures \cite{Bychkov1984:JETP}.

The band structure of graphene in the presence of transverse electric
field depends rather strongly on the interplay of the intrinsic and
extrinsic spin-orbit coupling effects. As the magnitude of the electric
field increases, we encounter the topologies on display in Fig. \ref{Fig:bands}.
If the BR splitting is lower than the intrinsic one, the spectral gap gets
smaller. The electron  branch of the spectrum is still
degenerate at $K(K')$; in contrast, the hole branch is split by
$2\lambda_{\rm BR}$. This topology can give a quantum
spin Hall insulator \cite{Kane2005:PRL}. Curiously, as the electric field is such that
$\lambda_{\rm BR} = \lambda_{\rm I}$, one of the hole branches rises,
forming a genuine touching cones structure of massless fermions with
one of the electron branches. The two remaining branches are parabolic
(massive). For fields such that $\lambda_{\rm BR} > \lambda_{\rm I}$, all
the branches are again parabolic, with a degeneracy of one electron and
one hole band. The calculated spectrum at $K(K')$  follows the
recipe $\mu\lambda_{\rm BR} + \nu|\lambda_{\rm BR}-\mu \lambda_{\rm I}|$, with
$\mu$ and $\nu$ being $\pm 1$.

The physics behind the calculated spectral topologies
can be described qualitatively by previously proposed effective Hamiltonians.
Without spin-orbit coupling the electronic band structure of graphene in the
vicinity of $K(K')$ is described by the Hamiltonian ${\cal
H}_0=\hbar v_{\rm F}(\kappa\sigma_x k_x+ \sigma_y k_y)$. Here $v_{\rm F}$
is the Fermi velocity, $k_x$ and $k_y$ are the cartesian components
of the electron wave vector measured from $K(K')$,
the parameter $\kappa = 1\,(-1)$ for the cones at $K\,(K')$, and $\sigma_x$
and $\sigma_y$ are the Pauli matrices acting on the so called pseudospin
space formed by the two triangular sublattices of graphene.
The Hamiltonian ${\cal H}_0$ describes gapless
states with conical dispersion $\varepsilon_0=\nu\hbar v_{\rm F}|\bm{k}|$
near the Dirac points. The eigenstates are $\psi_\nu=1/\sqrt{2}|\nu
e^{-i\kappa\varphi},\quad 1\rangle$ for the electron
band, $\nu=1$,  and hole band,  $\nu=-1$.

The intrinsic spin-orbit coupling is described by the effective
Hamiltonian ${\cal H}_{\rm
SO}=\lambda_{\rm I}\kappa\sigma_z s_z$ \cite{McClure1962,Kane2005:PRL}.
 Here $s_z$ is the spin Pauli
matrix. The spin-orbit coupling lifts the orbital degeneracy at
$K(K')$. Indeed, the eigenvalues of the combined Hamiltonian, ${\cal
H}_0+{\cal H}_{\rm SO}$, are
$\varepsilon_\nu=\nu\sqrt{\varepsilon_0^2+\lambda_{\rm I}^2}$. The
bands are split by $2|\lambda_{\rm I}|$, but the
two-fold degeneracy of the bands, required by space
inversion and time reversal symmetry, remains. The eigenvectors are
$\psi_{\mu\nu}=\chi_\mu |e^{-i\kappa\varphi}(\kappa\varepsilon_\nu+
\mu\lambda_{\rm I})/\varepsilon_0, \quad 1\rangle/C_{\mu\nu}$. Here
$\chi_\mu$ is the spin spinor with $\mu=\pm 1$; the normalization
constant is $C_{\mu\nu}=[1+(\kappa\varepsilon_\nu+\mu\lambda_{\rm
I})^2/ \varepsilon_0^2]^{1/2}$.

The extrinsic spin-orbit coupling of the Bychkov-Rashba type in
graphene can be described by the Hamiltonian
${\cal H}_{\rm BR}=\lambda_{\rm BR}(\kappa\sigma_x s_y-\sigma_y s_x)$,
where $\lambda_{\rm BR}$ is the Bychkov-Rashba parameter \cite{Kane2005:PRL}. Unlike in
semiconductor heterostructures, the coupling in graphene does not depend on the
magnitude of the electron momentum, as the electrons at  $K(K')$ have a constant velocity.
The electronic bands near $K\,(K')$ are now modified to
\begin{equation}\label{Eq:bands}
\varepsilon_{\mu\nu}=\mu\lambda_{\rm BR}+
\nu\sqrt{(\hbar v_{\rm F} k)^2+(\lambda_{\rm BR}-\mu\lambda_{\rm I})^2}\,.
\end{equation}
The corresponding eigenvectors are
\begin{eqnarray}
\psi_{\mu\nu}=(
\chi_- |\kappa e^{-i\kappa\varphi}[(\varepsilon_{\mu\nu}-\lambda_{\rm I})/
\varepsilon_0]^\kappa, \quad 1\rangle + \qquad\qquad
\\ \nonumber
\mu\chi_+ |-i\kappa e^{-i(1+\kappa)\varphi}, \quad
i e^{-i\varphi} [(\lambda_{\rm I}-\varepsilon_{\mu\nu})/\varepsilon_0]^\kappa
\rangle )/C_{\mu\nu}\,,
\end{eqnarray}
with the normalization constant
$C_{\mu\nu}=\sqrt{2}(1+[(\lambda_{\rm I}-\varepsilon_{\mu\nu})/
\varepsilon_0]^{2\kappa})^{/2}$. The expectation value of the spin,
\begin{equation}
s_{\mu\nu}= \frac{\varepsilon_0}{\sqrt{\varepsilon_0^2+(\lambda_{\rm
I}-\mu\lambda_{\rm BR})^2}} \left( \begin{array}{c} \sin\varphi \\
-\cos\varphi \\ 0 \end{array}\right),
\end{equation}
is $k$-dependent and lies in the graphene plane. The inclusion of
the extrinsic coupling lifts the two-fold degeneracy of the bands.
Only the time-reversal Kramers degeneracy remains, coupling states
at $K$ and $K'$.

Our first-principles results show that the above effective
Hamiltonian model gives a remarkably faithful description of graphene's
band structure at $K(K')$. The comparison is shown in Fig.
\ref{Fig:bands}. The dispersions given by Eq.~(\ref{Eq:bands})
differ from the numerical results by less than 5\% up to 
$\pm 200$~${\rm meV}$ away from the Fermi level.
With the parameters supplied by the first-principles calculations
the analytical model becomes highly accurate.

The extrinsic splitting $2\lambda_{\rm BR}$ is extracted as
$(\varepsilon_{+-}-\varepsilon_{--})/2$ for
$\lambda_{\rm BR}<\lambda_{\rm I}$, and $(\varepsilon_{++}-
\varepsilon_{--})/2$ for $\lambda_{\rm BR}>\lambda_{\rm I}$, 
at $K$. Figure ~\ref{Fig:efield}(a) illustrates the zigzag potential
modeling the transverse field. The calculated $2\lambda_{\rm BR}$
versus the electric field $E$ is shown in Fig.~\ref{Fig:efield}(b).
The dependence is linear with the slope of $10$~${\rm \mu\,eV\,nm/V}$. Since the
field of 1 V/nm is produced by an electron charge 1~${\rm nm}$
away from the graphene sheet, such fields are typical for graphene on
a substrate. We expect that in realistic situations the intrinsic and
extrinsic spin-orbit couplings compete, making the topologies
described in Fig.~\ref{Fig:bands} likely occurring in real samples.
Since the extrinsic coupling depends linearly on the electric field,
the topology is tunable by gates. We
also give the calculated magnitude of the graphene's dipole
moment (the shift of the electron charge density): 0.0134 ${\rm C\AA}$
in unit cell. One may then
relate the Bychkov-Rashba effect directly to the induced dipolar
moment; this should be particularly useful for estimating the extrinsic
splitting due to ad-atoms absorption on graphene.

Previous numerical estimates for the intrinsic splitting $2\lambda_{\rm I}$ in graphene
are rather controversial. The splitting was estimated to be in the range of 1 to 200~${\rm\mu eV}$
\cite{Kane2005:PRL,Min2006:PRB,Yao2007:PRB,Hernando2006:PRB,Boettger2007:PRB}.
Kane and Mele \cite{Kane2005:PRL} estimated the splitting of 200 $\mu$eV. This optimistic estimate was drastically reduced
by Min {\it et al.} \cite{Min2006:PRB} to the value of
1 $\mu$eV, supported by subsequent
works \cite{Yao2007:PRB, Hernando2006:PRB}. None of these studies
were fully first-principles. A density functional calculation of
Boettger and Trickey  \cite{Boettger2007:PRB}, using a
Gaussian-type orbital fitting function methodology, gave 
50 $\mu$eV. Our result is about one half of that; the difference is
likely due to the different approximation
schemes for spin-orbit coupling used in Ref. 
\onlinecite{Boettger2007:PRB} and by us \cite{endnote1}.
Previous estimates for the extrinsic splitting, $2\lambda_{\rm BR}$, are
0.516 $\mu$eV \cite{Kane2005:PRL} and 133 $\mu$eV \cite{Min2006:PRB} per V/nm.
No fully first-principles calculation of $2\lambda_{\rm BR}$, or the extrinsic
effects in graphene in general, was reported thus far.

What is the origin of the rather large, as compared to previous
non-fully-first-principles results, intrinsic spin-orbit
splitting in graphene? We calculate the spin-orbit coupling
splitting of the $2p$ levels in carbon atoms to be $8.74$~${\rm m\,eV}$,
using the \emph{Wien2k} code.
This splitting should be reflected in the splitting of the bands at the
$\Gamma$ point. Our calculation finds the splitting at the $\Gamma$
point of $8.978$~${\rm m\,eV}$ about $3$~eV below the Fermi level,
in close agreement with the atomic value. The bands at the $K\,(K')$
points are formed mainly by $p_z$ orbitals whose magnetic quantum
number is zero. The intrinsic splitting can be due to the coupling of the $p_z$ orbitals
(forming the $\pi$-bands) to either $\sigma$ bands or bands formed
by higher orbitals ($d$, $f$, ...). As argued already
by Slonczewski \cite{Slonczewski:thesis} using group theory,
it is the $d$ and higher orbitals that dominate the spin-orbit
splitting at $K\,(K')$. A qualitative argument for that was
provided by McClure and Yafet \cite{McClure1962}: Orbitals $d_{xz}$
and $d_{yz}$ can form Bloch states of the $\pi$ band symmetry at $K(K')$.
Due to a finite overlap between the neighboring $p_z$ and $d_{xz}, d_{yz}$
orbitals, the intrinsic splitting is linearly proportional to the spin-orbit splitting
of the $d$ states (orbitals higher than $d$ have a smaller overlap and contribute
less). In contrast, due to the absence of the direct overlap between the
$p_z$ and $\sigma$-band orbitals, the usually considered \cite{Kane2005:PRL,Min2006:PRB,Yao2007:PRB,Hernando2006:PRB}
spin-orbit splitting induced by the $\sigma$-$\pi$ mixing
depends only quadratically on the atomic spin-orbit splitting,
giving a negligible contribution.

In Fig.~\ref{Fig:efield}(c) we show the orbital-resolved densities of states.
The $p_x, p_y$ atomic character vanishes above $-3\,{\rm eV}$ (the Fermi
 level $\varepsilon_{\rm F}$ is at zero). The density of states close to
$\varepsilon_{\rm F}$ comes predominantly from $p_z$ orbitals ($\pi$-bands).
Nevertheless, there is a finite contribution from $d$ orbitals ($d_{xz}$ and $d_{yz}$) which
follows in shape that of $p_z$. The contributions from $d_{x^2-y^2}$, $d_{xy}$, and $d_{z^2}$
vanish for energies above the $-3\,{\rm eV}$.
To further confirm the symmetry arguments of Slonczewski, we have
selectively removed orbitals from the calculation of the spin-orbit
coupling contribution. Removing $d$ and higher orbitals reduces the
intrinsic splitting to 0.98 ${\rm \mu\,eV}$, reproducing earlier non-first-principles
calculations \cite{Min2006:PRB,Yao2007:PRB,Hernando2006:PRB}. We
conclude that the intrinsic splitting is dominated by $d$
and higher orbitals, giving more that 96\% of the splitting
\cite{endnote2}.
In contrast, we find that the extrinsic BR coupling is largely unaffected by the presence of $d$
and higher orbitals, demonstrating that this coupling is due to the
$\sigma$-$\pi$ mixing.

Finally, we calculate the spin-orbit splitting for what we call
a `mini-ripple' configuration (Fig.~\ref{Fig:miniripple}), in
which every other atom is displaced by 
an amount $\Delta$ transverse to the sheet. This
calculation should give an indication of what to expect for larger-scale
ripples that occur in graphene on a substrate \cite{Ishigami2007:NL}
or free standing \cite{Fasolino2007:NM}; such
large scales are out of the scope for our methods.
The mini-ripple exhibits a gap that grows quadratically
with increasing $\Delta$, as seen from Fig.~\ref{Fig:miniripple}.
Removing $d$ and higher orbitals from the calculation of the
spin-orbit splitting, the initial gap reduces to about 1$\,{\rm \mu\,eV}$
but the overall growth remains largely unchanged: the rippling-induced
gap is almost solely due to $\sigma$-$\pi$ mixing. Furthermore,
the displacements of less than 1\% have no significant effects 
on the intrinsic spin-orbit splitting. For an effective description 
of the gap opening in the mini-ripple we consider a more general 
extrinsic case. The contribution of a transverse electric field 
is twofold. First, there is a pseudospin splitting, as the two
sublattices are no longer equivalent (the two triangular sublattices
have a different potential). 
Second, the Bychkov-Rashba effect appears. We find that the 
following formula describes the resulting spectrum:
$\epsilon_{\mu\nu}^K=\nu\lambda_{\rm I} + \mu[\delta_{1\nu}
\lambda_\Delta +\delta_{-1\nu}\sqrt{\lambda_\Delta^2+(2\lambda_{\rm
BR})^2}]$, where $\lambda_\Delta$ is the electric field induced
gap due to the rippling ($\lambda_\Delta \propto E$).

In summary, we have shown that for realistic electric fields the
graphene band structure exhibits remarkable tunable topologies.
Our first-principles calculation gives strong support for
the effective spin-orbit coupling Hamiltonian models, making them
highly accurate analytical tools to investigate the physics of
graphene.

We thank A. Matos-Abiague, P. Blaha, S. B. Trickey, and the {\it Wien2k} community for useful hints and discussions. This work was supported by the DFG SFB 689 and SPP 1285. CAD appreciates
support from the Austrian Science Fund, Project I107.


\begin{thebibliography}{10}

\bibitem{Novoselov2004:Science}
K.~S. Novoselov, {\it et~al.\/}, {\it Science\/} {\bf 306}, 666 (2004).

\bibitem{Novoselov2005:Nature}
K.~S. Novoselov, {\it et~al.\/}, {\it Nature\/} {\bf 438}, 197 (2005).

\bibitem{Zhang2005:Nature}
Y.~Zhang, Y.-W. Tan, H.~L. Stormer, P.~Kim, {\it Nature\/} {\bf 438}, 201
  (2005).

\bibitem{Bostwick2007:NP}
A.~Bostwick, T.~Ohta, T.~Seyller, K.~Horn, E.~Rotenberg, {\it Nature Phys.\/}
  {\bf 3}, 36 (2007).

\bibitem{Geim2007:NM}
A.~K. Geim, K.~S. Novoselov, {\it Nature Mat.\/} {\bf 6}, 183 (2007).

\bibitem{Adam2007:PNAS}
S.~Adam, E.~H. Hwang, V.~M. Galitski, S.~{Das Sarma}, {\it PNAS\/} {\bf 104},
  18392 (2007).

\bibitem{Tombros2007:Nature}
N.~Tombros, C.~Jozsa, M.~Popinciuc, H.~T. Jonkman, B.~J. van Wees, {\it
  Nature\/} {\bf 448}, 571 (2007).

\bibitem{Kane2005:PRL}
C.~L. Kane, E.~J. Mele, {\it Phys. Rev. Lett.\/} {\bf 95}, 226801 (2005).

\bibitem{Trauzettel2007:NP}
B.~Trauzettel, D.~V. Bulaev, D.~Loss, G.~Burkard, {\it Nature Phys.\/} {\bf 3},
  192 (2007).

\bibitem{Wallace1947:PR}
P.~R. Wallace, {\it Phys. Rev.\/} {\bf 71}, 622 (1947).

\bibitem{Coulson1952:PPS}
C.~A. Coulson, R.~Taylor, {\it Proc. Phys. Soc. A\/} {\bf 65}, 815 (1952).

\bibitem{Lomer1955:PRS}
W.~M. Lomer, {\it Proc. Roy. Soc. (London)\/} {\bf A227}, 330 (1955).

\bibitem{McClure1957:PR}
J.~W. McClure, {\it Phys. Rev.\/} {\bf 108}, 612 (1957).

\bibitem{Slonczewski1958:PR}
J.~C. Slonczewski, P.~R. Weiss, {\it Phys. Rev.\/} {\bf 109}, 272 (1958).

\bibitem{McClure1962}
J.~W. McClure, Y.~Yafet, {\it Proceedings of the Fifth Conference on Carbon\/}
  (Pergamon, New York, 1962), vol.~1, pp. 22--28.

\bibitem{Slonczewski:thesis}
J.~C. Slonczewski, Band theory of graphite, Ph.D. thesis, Rutgers University of
  New Jersey (1955).

\bibitem{Blaha:Wien2k}
P.~Blaha, K.~Schwarz, G.~Madsen, D.~Kvasnicka, J.~Luitz, {\it WIEN2k - An
  Augmented Plane Wave Plus Local Orbitals Program for Calculating Crystal
  Properties\/}, TU Wien, Austria (2008).

\bibitem{Perdew1996:PRL}
J.~P. Perdew, K.~Burke, M.~Ernzerhof, {\it Phys. Rev. Lett.\/} {\bf 77}, 3865
  (1996).

\bibitem{Singh2005:book}
D.~J. Singh, L.~Nordstrom, {\it Planewaves, Pseudopotentials, and LAPW
  method\/} (Springer, 2005).

\bibitem{Stahl2001:PRB}
J.~Stahl, U.~Pietsch, P.~Blaha, K.~Schwarz, {\it Phys. Rev. B\/} {\bf 63},
  165205 (2001).

\bibitem{Fabian2007:APS}
J.~Fabian, A.~Matos-Abiague, C.~Ertler, P.~Stano, I.~\v{Z}uti\'c, {\it Acta
  Phys. Slov.\/} {\bf 57}, 565 (2007).

\bibitem{Bychkov1984:JETP}
Y.~A. Bychkov, E.~I. Rashba, {\it JETP Lett.\/} {\bf 39}, 78 (1984).

\bibitem{Min2006:PRB}
H.~Min, {\it et~al.\/}, {\it Phys. Rev. B\/} {\bf 74}, 165310 (2006).

\bibitem{Yao2007:PRB}
Y.~Yao, F.~Ye, X.-L. Qi, S.-C. Zhang, Z.~Fang, {\it Phys. Rev. B\/} {\bf 75},
  041401 (2007).

\bibitem{Hernando2006:PRB}
D.~Huertas-Hernando, F.~Guinea, A.~Brataas, {\it Phys. Rev. B\/} {\bf 74},
  155426 (2006).

\bibitem{Boettger2007:PRB}
J.~C. Boettger, S.~B. Trickey, {\it Phys. Rev. B\/} {\bf 75}, 121402 (2007).

\bibitem{Ishigami2007:NL}
M.~Ishigami, J.~H. Chen, W.~G. Cullen, M.~S. Fuhrer, E.~D. Williams, {\it Nano
  Lett.\/} {\bf 7}, 1643 (2007).

\bibitem{Fasolino2007:NM}
A.~Fasolino, J.~H. Los, M.~I. Katsnelson, {\it Nature Mat.\/} {\bf 6}, 858
  (2007).

\bibitem{endnote1}{The precise value of the intrinsic gap depends on the muffin-tin radius, since \protect \emph  {Wien2k} calculates spin-orbit coupling only inside the muffin-tin sphere. (The present calculations use the maximum allowed sphere radius.) Although spin-orbit coupling is concentrated around the atom cores, making the inner-sphere approximation rather accurate, it is likely that the actual value of the gap is somewhat larger.}
\bibitem{endnote2}{The contribution to $\lambda _{\protect \rm  I}$ changes nonmonotonic as the maximum orbital state $\ell _{\protect \rm  max}$ in the second variational scheme is increased. For $\ell _{\protect \rm  max}=1$, 2, and 3 and more, we get $2\lambda _{\protect \rm  I}\approx 1$, 26, and 24 $\mu $eV.}

\end{thebibliography}


\begin{figure*}
\centering
\includegraphics[width=0.60\columnwidth,angle=0]{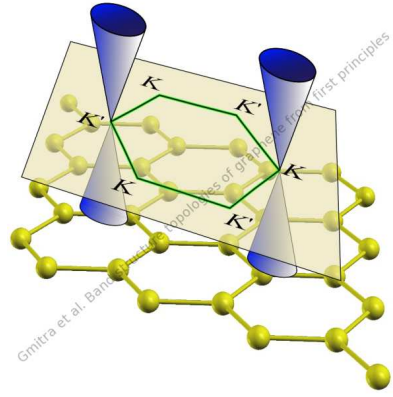}
\caption{Graphene's essentials. Bottom-up: Carbon atoms form a honeycomb lattice with
two atoms in the unit cell. The first Brillouin zone of the reciprocal lattice contains two
nonequivalent Dirac points, $K$ and $K'$. The relevant states at the Fermi level form two 
touching cones with the tips at $K(K')$.}
\label{Fig:sketch}
\end{figure*}

\begin{figure*}
\centering
\includegraphics[width=2.0\columnwidth,angle=0]{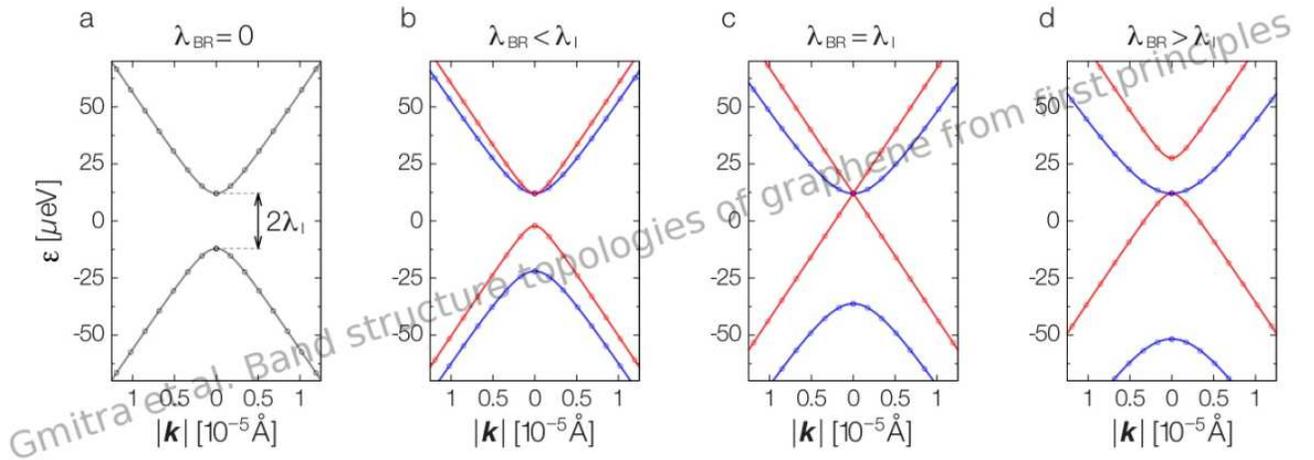}
\caption{Band structure topologies of graphene.
Transverse electric
field drastically changes the topology of the bands near $K(K')$.
(\textbf{a})~Zero electric field. (\textbf{b})~Electric field of magnitude
$E=1.0\,{\rm V/nm}$. (\textbf{c})~$E=2.44\,{\rm V/nm}$.
(\textbf{d})~$E=4.0\,{\rm V/nm}$.
The first-principles results are represented by circles (the Fermi level
is at zero). The curves are
fits to the analytical model (see the main text).
The spin branch $\mu=1$ is shown in red, $\mu=-1$ in blue. The calculated
Fermi velocity is $v_{\rm F}=0.833\times 10^6\,{\rm m/s}$.
}
\label{Fig:bands}
\end{figure*}

\begin{figure*}
\centering
\includegraphics[width=2.0\columnwidth,angle=0]{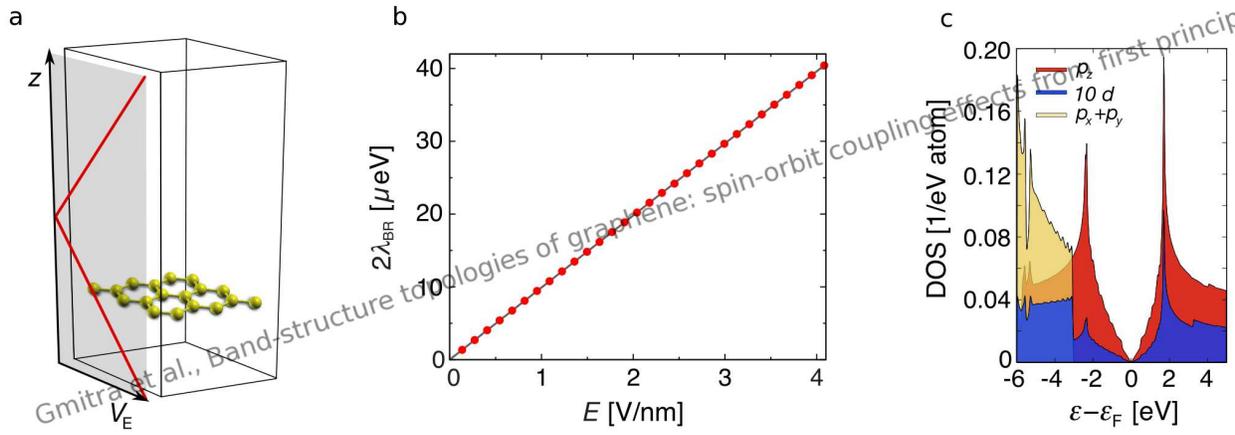}
\caption{Bychkov-Rashba-type splitting in graphene and orbital-resolved
density of states. (\textbf{a})~Transverse electric field is modeled with a zigzag potential. 
(\textbf{b})~Bychkov-Rashba spin-orbit induced splitting at $K(K')$
as a function of the electric field. The slope is 9.9 $\mu$eV per V/nm.
(\textbf{c})~Projected density of states to particular atomic
orbitals. The Fermi level is at zero. The $d$-character is enhanced by a factor of ten.
}
\label{Fig:efield}
\end{figure*}

\begin{figure*}
\centering
\includegraphics[width=1.0\columnwidth,angle=0]{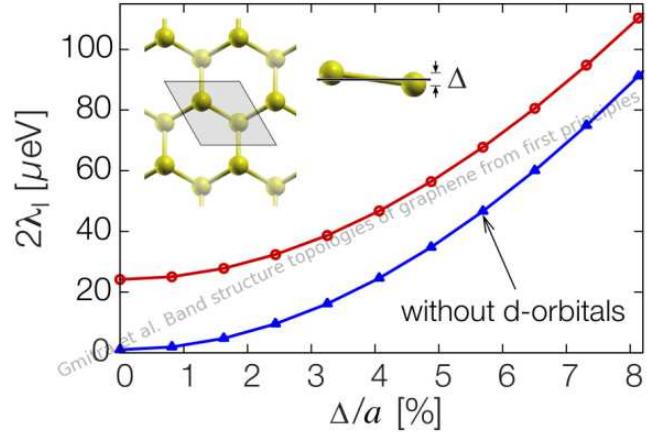}
\caption{Calculated spectral gap at $K(K')$ in the mini-ripple configuration (inset),
as a function of the relative corrugation strain with respect to
the lattice constant $a$. Neglecting $d$ and higher orbitals
results in an almost constant shift, proving that the main effects come
from $\sigma$-$\pi$ hybridization.
}
\label{Fig:miniripple}
\end{figure*}

\end{document}